\documentclass[amsmath,amssymb,superscriptaddress,10pt,nofootinbib,reprint,notitlepage]{revtex4-1}

\usepackage[utf8]{inputenc}

\usepackage{graphicx} 
\usepackage{bm}
\usepackage{amsmath}
\usepackage{amssymb}
\usepackage{mathrsfs}
\usepackage{mathtools}
\usepackage{url}

\usepackage{textcomp} 

\usepackage[colorlinks,linkcolor=blue,citecolor=blue,urlcolor=blue]{hyperref}
\usepackage{xcolor}

\usepackage{verbatim}
\usepackage{color}

\usepackage{float}

\usepackage{tikz}                       
\usepackage{pgfplots}                   
\pgfplotsset{compat=1.9}
\usetikzlibrary{calc,intersections}
\usepgfplotslibrary{external}
\usetikzlibrary{pgfplots.groupplots}
\usepgfplotslibrary{fillbetween}

\begin{document}

\title{A Primary Radiation Standard Based on Quantum Nonlinear Optics }

\author{Samuel~Lemieux}
\email[]{samzlemieux@gmail.com}
\affiliation{Department of Physics and Max Planck Centre for Extreme and Quantum Photonics, University of Ottawa, 25 Templeton Street, Ottawa, Ontario K1N 6N5, Canada}

\author{Enno~Giese}
\affiliation{Institut für Quantenphysik and Center for Integrated Quantum Science and Technology $\left(\text{IQ}^{\text{ST}}\right)$, Universität Ulm, Albert-Einstein-Allee 11, D-89081, Germany}

\author{Robert~Fickler}
\affiliation{Institute for Quantum Optics and Quantum Information (IQOQI), Austrian Academy of Sciences, Boltzmanngasse 3, 1090 Vienna, Austria}

\author{Maria~V.~Chekhova}
\affiliation{Max Planck Institute for the Science of Light, G.-Scharowsky Str.1/Bau 24, 91058 Erlangen, Germany}
\affiliation{Physics Department, Lomonosov Moscow State University, Moscow 119991, Russia}
\affiliation{University of Erlangen-Nuremberg, Staudtstrasse 7/B2, 91058 Erlangen, Germany}

\author{Robert~W.~Boyd}
\affiliation{Department of Physics and Max Planck Centre for Extreme and Quantum Photonics, University of Ottawa, 25 Templeton Street, Ottawa, Ontario K1N 6N5, Canada}
\affiliation{Institute of Optics, University of Rochester, Rochester, New York 14627, USA}

\date{\today}

\begin{abstract}
The spectrum of vacuum fluctuations of the electromagnetic field is determined solely from first physical principles and can be seen as a fundamental property that qualifies as a primary radiation standard.
We demonstrate that the amplitude of these quantum fluctuations triggering nonlinear optical processes can be used as a reference for radiometry.
In the spontaneous regime of photon pair generation, the shape of the emitted spectrum is nearly independent of laboratory parameters.
In the high-gain regime, where spontaneous emission turns to stimulated emission, the shape of the frequency spectrum is uniquely determined by the number of created photons.
Both aspects allow us to determine the quantum efficiency of a spectrometer over a broad range of wavelengths without the need of any external calibrated source or detector. 
\end{abstract}

\maketitle
\normalsize

\noindent
The desire to understand thermal radiation helped lead to the development of quantum mechanics.
For its part, quantum mechanics was crucial for the accurate description of electromagnetic radiation. 
As a consequence, the black body\textemdash a perfect absorber at thermal equilibrium\textemdash remains to this day \emph{the} primary source of light for radiometry~\cite{hollandt20055}.
Currently, the only alternative is synchrotron radiation, whose description relies on classical electrodynamics and which requires costly and large facilities~\cite{lemke1967synchrotron}.
In our article, we exploit the quantum properties of nonlinear optical processes to introduce a primary radiometric standard that is straightforwardly realized with equipment available in most quantum optics labs. 

The quantum-mechanical fluctuations of the electromagnetic vacuum, originating from the non-vanishing bosonic commutation relation of the photons, exhibit a unique frequency spectrum. 
At the same time, the rate of spontaneous photon generation crucially depends on the amplitude of those fluctuations.   
We use the spectrum of the vacuum, as well as its nonlinear amplification, as a primary standard to infer the spectral response and efficiency of an optical system.
Parametric down-conversion~(PDC), a nonlinear optical process based on three-wave mixing with only one input field, gives us direct access to the bare spectrum of the vacuum. 
In fact, the vacuum fluctuations are the dominant frequency-dependent contribution to phase-matched spontaneous PDC, resulting in a spectral shape that is nearly independent of any laboratory parameters. 
In the high-gain regime, the nonlinear amplification distorts this spectrum in a specific way, allowing one to extract the number of down-converted photons only from the spectral shape of the emission. 
Thus, the spectrum of the vacuum fluctuations that leads to the creation of the biphoton field is a standard that gives us access to a radiometric realization, which is here the number of emitted photons.  

Photon-pair generation lies at the heart of other radiometric calibration methods.
Coincidence measurements, for instance, have been used to determine the quantum efficiency of photodetectors~\cite{klyshko1980use,malygin1981absolute,polyakov2007high,rarity1987absolute}.
In another strategy, which also relies on the brightness of the vacuum, the ratio between seeded and unseeded PDC allows one to measure the spectral radiance of a light source~\cite{klyshko1977utilization,kitaeva1979measurement,migdall1998measuring}.
However, in contrast to the latter, our method exploits the spectrum of the vacuum itself, as well as its unique behaviour in the strong-coupling regime. 
In that respect, the same radiometric principles pertaining to black-body radiation can be applied to our source. 
Using these insights, we determine the absolute quantum efficiency of a spectrometer over a broad spectral range, without using any reference detector.  
As a first step, we obtain the spectral response of the spectrometer using spontaneous PDC\textemdash this is a relative calibration. 
Then, we deduce the parametric gain and the spectral quantum efficiency from the shape of high-gain PDC spectra. 
Our experimental results compare well against the ones obtained with a reference lamp, and the quantum efficiency agrees with expected values, thereby demonstrating a promising novel method to produce a primary radiation standard. 

In general, a source can serve as a primary radiation standard if, within a specified bandwidth centered on the wavelength $\lambda$, the exact number of emitted photons $N(\lambda)$ is known. 
However, the number of counts $M(\lambda)$ recorded by a detector do not, usually, coincide with $N(\lambda)$ due to a non-perfect quantum efficiency $\eta(\lambda)$ of the detecting device.
These quantities are simply connected through the relation
\begin{equation}
M(\lambda) = \eta(\lambda) \, N(\lambda).
\label{eq:1}
\end{equation}
Measuring $M(\lambda)$ while having a precise knowledge of $N(\lambda)$ allows the determination of $\eta(\lambda)$, which is at the heart of the absolute calibration of spectrometers.
The spectral efficiency $\eta(\lambda)$ can be further separated into its relative spectral shape $R(\lambda)$, i.e. the response function of the measurement device, and a wavelength-independent proportionality constant $\alpha$\textemdash through $\eta(\lambda) = \alpha R(\lambda)$. 
While a relative calibration procedure gives $R(\lambda)$, obtaining the full spectral quantum efficiency $\eta(\lambda)$ requires an absolute calibration. 
In the following we demonstrate in a two-step procedure that both relative and absolute calibration can be performed using PDC.

The total number of photons $N(\lambda)$  reaching the detector depends on the photon-number distribution $\mathcal{N}$ per plane-wave mode characterizing the source, and on the modes that are detected.
Using standard radiometric formalism, this fact translates to the expression~\cite{suppl}
\begin{align}
\begin{split}
N(\lambda) &= \frac{1}{(2 \pi)^3} \smashoperator{\int_{\textrm{source}} } \textrm{d}^3 r \smashoperator{\int_{\text{detector}}} \textrm{d}^3 k \, \mathcal{N} \\
 &\approx [A_s \, c \, \tau_s ]\,[\Delta \Omega \Delta \lambda]\, \mathcal{D}(\lambda) \, \mathcal{N}, 
\label{eq:2}
\end{split}
\end{align} 
where the first integral can be approximated by the transverse area $A_s$ of the source and the duration of the emission $\tau_s$ multiplied by the speed of light $c$.
The second integral incorporates the modes that are detected and can be approximated by the bandwidth $\Delta \lambda$ and solid-angle  $\Delta \Omega$ of the detector, if $\mathcal{N}$ does not vary significantly over these quantities. 
In order to connect the plane waves to the solid angle and the wavelength, which are the relevant quantities for a spectrometer~\footnote{In radiometry, the quantity $c\, \mathcal{D}(\lambda) (2\pi)^{-3} \, \mathcal{N} (h c / \lambda )$ is the spectral radiance, or the energy per units of time, area of the source, solid angle and bandwidth (in wavelength) of the detector.}, we also introduced the quantity $\mathcal{D}(\lambda) = (2\pi)^3\, \lambda ^{-4}$, which is proportional to the mode density~\cite{klyshko1989photons}.
If $\mathcal{N}$ is known, we have all the necessary quantities for the absolute calibration of a spectrometer.
For black-body radiation, $\mathcal{N}$ is derived from physical principles, namely the photon-distribution at thermal equilibrium with a certain temperature.   

\begin{figure}[htb] \centering
\includegraphics{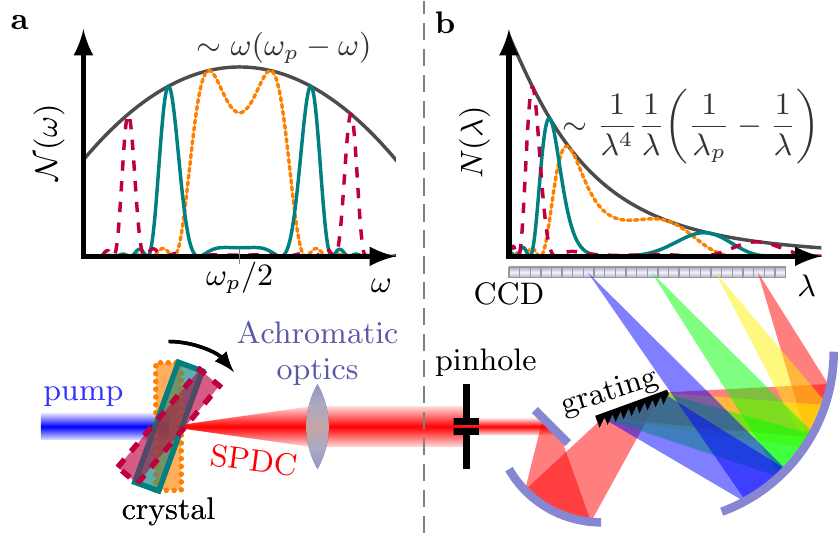}
\caption{
Physical principle and idealized setup. 
The shape of the phase-matched photon-number distribution $\mathcal{N}_\text{PM}$ in a given direction depends on the amplitude of vacuum fluctuations.
a) Different tilt angles for the nonlinear crystal correspond to different phase-matching conditions, altering the spectrum $\mathcal{N}$ accordingly.
$\mathcal{N}_\text{PM}$ (solid black) is obtained by taking the maximum of $\mathcal{N}$ for different phase-matching conditions. 
 b) The photon-number spectrum $N(\lambda)$ is measured with an angular filter (a pinhole in the far field selects a small solid angle) and a spectrometer. 
The additional $\lambda^{-4}$ factor in $N(\lambda)$ relates the plane-wave representation to the basis of the spectrometer. 
}
\label{fig:principle}
\end{figure}

During the three-wave mixing process of PDC, pump photons (of frequency $\omega_p$) interact with the vacuum field within a crystal with a $\chi^{(2)}$ nonlinearity.
This process leads to the generation of pairs of photons known as the signal and the idler. 
In the spontaneous regime~(low pump intensity), the photon distribution $\mathcal{N}$, a function of frequency and emission angle, depends on the amplitude of the vacuum fluctuations, the profile of the pump beam, the gain of the amplification process and a phase-matching function.
For a monochromatic plane wave pump of amplitude $E_p$ and a crystal of thickness $L$, the photon-number distribution of spontaneous PDC is given by
\begin{equation}
\mathcal{N} =   \left(c^{-1}\, L\, \chi ^{(2)} E_p\right)^2 \sqrt{ \omega \omega_i/(n n_i)}^2\operatorname{sinc} ^2( \Delta \kappa L/2)
\label{eq:n}
\end{equation}
where $n$ and $n_i$ are the signal and idler refractive indices, and $\Delta \kappa = \kappa_p - \kappa - \kappa_i$ is the mismatch between the longitudinal wave vectors of the pump, the signal, and the idler, respectively~\cite{klyshko1989photons,lemieux2016engineering}.
The frequency-dependent factors $\sqrt{\omega/n}$ and $\sqrt{\omega_i /n_i}$ arise from the quantization of the electric field for the signal and for the idler.
In the spontaneous regime of pair creation, those factors embody the amplitude of the vacuum fluctuations for the biphoton field. 
To denote the coupling strength, we use the gain parameter $\mathcal{G} = c^{-1}\, L\, \chi^{(2)} E_p / \sqrt{n\, n_i}$, which we can assume to be constant over the frequency range of interest~\cite{suppl}.
 
The last factor of equation~(\ref{eq:n}) is the well-known phase-matching function of a bulk crystal. 
At exact phase-matching, $\Delta \kappa$ vanishes and the phase-matching function takes on the value unity.
Thus, the phase-matched distribution takes its maximal value and reads
\begin{equation}
\mathcal{N}_\text{PM} = \mathcal{G}^2\, \omega(\omega_p - \omega),
\label{eq:pm}
\end{equation}
where we assumed that photon energy is conserved in the parametric process, such that $\omega_i = \omega_p - \omega$.
For absolute calibration, we need a complete knowledge of $\mathcal{N}_\text{PM}$, but it is difficult to determine $\mathcal{G}$ experimentally in the spontaneous regime of PDC.
However, we note that the photon number for different phase-matching conditions $\mathcal{N}_\text{PM}$ follows a parabola, as illustrated in Fig.~\ref{fig:principle}a.
Because $\omega(\omega_p - \omega)$ does not depend on laboratory parameters, we can make use of the shape of $\mathcal{N}_\text{PM}$ and perform a relative calibration~\cite{[See supplementary material of ]lemieux2016engineering}.

By introducing a pinhole in the far-field of the crystal, we can limit the emission solid-angle, thereby suppressing the frequency content in the other angular modes~\cite{suppl}.
A spectrometer then disperses the light and images it onto a CCD chip; see Fig.~\ref{fig:principle}b.
Since the position on the chip corresponds to a particular wavelength, we expect a specific functional behavior that originates in the parabola but is modified by the different mode density $\mathcal{D}(\lambda)$~\cite{spasibko2012spectral}.
However, any deviation of the phase-matched wavelength from this function can be assigned to detector inefficiencies and therefore to $R(\lambda)$.

Since $R(\lambda)$ is proportional to the ratio between $M(\lambda)$ and the shape of $N(\lambda)$, we can write 
\begin{equation}
R(\lambda) \propto \frac{M(\lambda)}{\mathcal{D}(\lambda) \omega (\omega_p -\omega)}\bigg\rvert_{\textrm{PM}},
\label{eq:r}
\end{equation}
where $\omega = 2\pi c /\lambda$ and where we used the proportionality symbol because $\mathcal{G}$ has yet to be determined. 
The right-hand side is evaluated at the wavelength $\lambda_\textrm{PM}$ that satisfies the phase-matching condition.

\begin{figure} \centering
\includegraphics{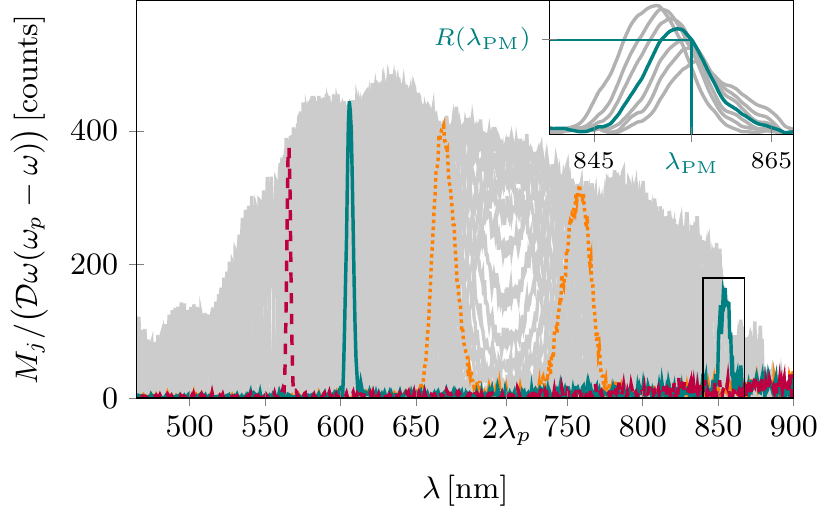}
\caption{Extracting the response function $R(\lambda)$ from the overlap of 411 measured spectra, in gray. 
The twin-peak structure in the orange-dotted and teal-solid spectra is a feature of phase matching and energy conservation.
For the red-dashed curve, the second peak does not lie within our measurement range. 
The maximum possible signal at a certain wavelength $\lambda$ is proportional to $R(\lambda)$. 
To illustrate this method, the inset shows several spectra (Fourier-filtered to suppress the noise) from the box enclosing the right-hand peak of the teal curve. 
 }
\label{fig:singlespectra}
\end{figure}

In our experiment, we pump a BBO crystal with a pulsed laser of wavelength $355\,\textrm{nm}$ and acquire a large number of spectra $M_j$ corresponding to different phase-matching conditions over a broad spectral range.
The phase-matched wavelengths are tuned by tilting the nonlinear crystal, as shown in the bottom panel of Fig.~\ref{fig:principle}a.  
We overlap all the measured spectra in Fig.~\ref{fig:singlespectra} and highlight three of them to show their twin-peak structure.
Importantly, the peak number of counts in a measured spectrum does not always occur at $\lambda_\textrm{PM}$. 
In fact, any nonzero slope to $R(\lambda)$ displaces the peak and a simple analysis shows that it shifts in the order of $\Lambda \approx (\textrm{d}R /\textrm{d}\lambda) (\sigma_\lambda ^2 / R) $, where $\sigma_\lambda$ is the standard deviation of the phase-matching function, approximated by a Gaussian. 
We see that the displacement from $\lambda_\text{PM}$ increases for a steep $R$ and a wide phase-matching function.
Hence, $\lambda_\text{PM}$ can differ significantly from the wavelength of the peak.
However, when overlapping different spectra, the maximum number of counts at one particular wavelength always yields the phase-matched measurement $M_j(\lambda_\textrm{PM})$ which follows directly from the fact that $\operatorname{sinc}^2 (\Delta \kappa L/2) \leq 1$.
We show this effect in the inset of Fig.~\ref{fig:singlespectra} and give more details on the data analysis in the supplementary material~\cite{suppl}.

 \begin{figure}[htb]  \centering
\includegraphics{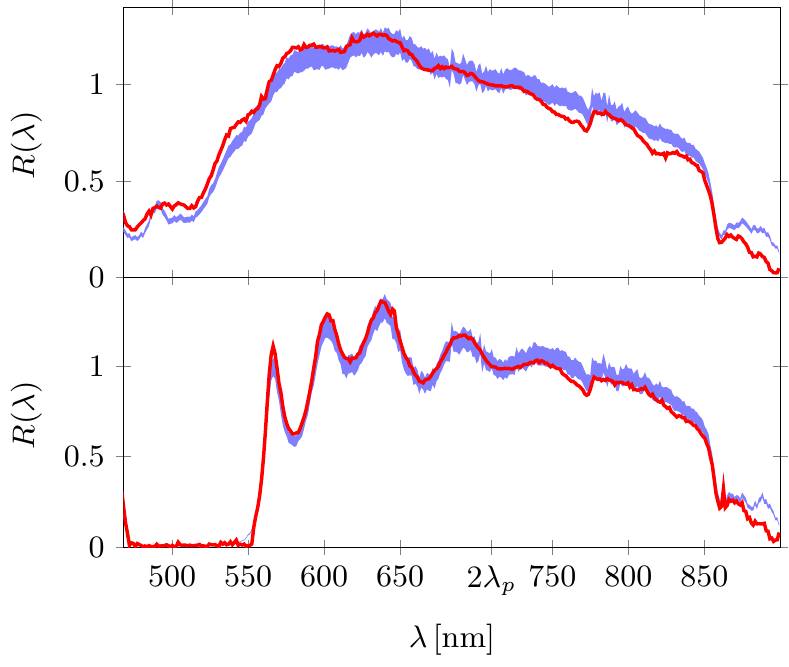}
\caption{
Comparison of the response function $R(\lambda)$ obtained from spontaneous PDC (solid, red, normalized to unity at $2 \lambda_p$) and the response function measured with a reference lamp (blue region enclosing the 5\% error reported by the manufacturer; scaled onto the PDC curves using a linear fit). 
To obtain the curves in the bottom panel, we added a dichroic filter to the spectrometer to induce rich spectral features into the response function. 
}
\label{fig:resp}
\end{figure}

We perform the experiment in the spontaneous regime of PDC, where the number of generated photons scales linearly with the pump intensity, in order to ensure the validity of equation~\eqref{eq:n}.
We retrieve $R(\lambda)$ directly from the spectra by virtue of equation~\eqref{eq:r}, where we find $\lambda_{\textrm{PM}}$ by taking the maximum of many spectra and we apply the arbitrary normalization $R(2\lambda_p) = 1$, such that $\eta(2\lambda_p) = \alpha$.
The response function obtained from the spontaneous PDC agrees very well with the response function measured with a reference lamp (Fig.~\ref{fig:resp}).
The experiment was repeated with an additional dichroic filter to demonstrate that the method resolves rich and rapidly varying spectral features. 
For a proper comparison, it is crucial that the light from spontaneous PDC and from the reference lamp undergo the exact same transfer function.
Thus, deviations stem from chromatic aberration, non-perfect polarization filtering as well as inaccuracies in the reference spectrum of the lamp.

To improve the precision of our method, one could include the frequency dependence of $\mathcal{G}$ if the linear and nonlinear dispersion relations of the crystal are known.
In this case, it is also straightforward to generalize equation~(\ref{eq:n}) so that it incorporates the spatial and temporal profiles of the pump beam~\cite{hsu2013absolute}.
Since we want to stress the simplicity of our procedure, we refrain from applying these corrections, but nonetheless obtain excellent results. 
With a knowledge of $R(\lambda)$\textemdash the form of $\eta(\lambda)$\textemdash we can accurately measure the \emph{shape} of any spectrum.
In the following, we perform the second step of our calibration procedure, the establishment of an absolute calibration method. 
In particular, we extract the number of photons from the shape of high-gain PDC spectra, based on our previous measurement of $R$. 

For an arbitrary value of the gain, the photon-number distribution under phase matching and for a monochromatic, plane-wave undepleted pump, becomes 
\begin{equation}
\mathcal{N}_\text{PM} =  \sinh ^2 \left( \mathcal{G}  \sqrt{\omega (\omega_p - \omega) }\right),
\label{eq:hgpm}
\end{equation}
which reduces to equation~\eqref{eq:pm} in the spontaneous regime, i.e., for $\mathcal{G}~\ll~1$~\cite{klyshko1989photons}. 
In the high-gain regime, the phase-matched photon-number spectrum is therefore a distorted parabola, whose spectral shape (curvature) and photon number are uniquely determined by the gain parameter $\mathcal{G}$.  
In complete analogy to equation~\eqref{eq:r} we obtain the relation 
\begin{equation}
\alpha  \sinh ^2 \left( \mathcal{G} \sqrt{\omega (\omega_p - \omega) }  \right) 
= \frac{M(\lambda) }{R(\lambda) \mathcal{D}(\lambda) \, \Gamma}\bigg\rvert_{\textrm{PM}}, 
\label{eq:fit}
 \end{equation} 
where we introduced, for a more convenient notation, the constant $\Gamma = \Delta \Omega  \Delta \lambda \, A_s  c \tau_s$ for the emission and detection parameters. 
Note that, in contrast to equation~\eqref{eq:r}, we now have an equality.
Except for $\alpha$, all the quantities are known: we obtained $R(\lambda)$ from spontaneous PDC, and the shape of the phase-matched spectrum uniquely determines $\mathcal{G}$.
We approximate $A_s$ by the transverse area of the pump beam, and $\tau_s $ by $ m \tau_p$, with $\tau_p$ being the pump pulse duration and $m$ the number of pulses during an acquisition time. 
Further, we calculate the solid angle $\Delta \Omega$ from the pinhole size in the far field of the crystal, and obtain $\Delta \lambda$ from the bandwidth associated with a pixel of the spectrometer's camera.
The only remaining free parameter, $\alpha$, is obtained via fitting. 

The experimental procedure for absolute calibration with high-gain PDC is very similar to the one for spontaneous PDC.
We acquire a large number of densely packed spectra $M_j(\lambda)$ for different crystal tilt angles with a much higher pump energy per pulse to reach a large parametric gain.
After taking the maxima of these dense spectra, we perform a bivariate curve fit using the free parameter $\alpha$ and the pump-normalized gain $\mathcal{G} / E_p$, a quantity that allows us to suppress the pulse energy drift of our pump laser over the acquisition time, and where a \emph{relative} measurement of $E_p$ is sufficient. 
We then obtain the spectral quantum efficiency by taking the product $\eta (\lambda) = \alpha  R(\lambda)$, with $R$ inferred from the spontaneous measurement and $\alpha$ from the high-gain regime. 

\begin{figure}[htb] \centering
\includegraphics{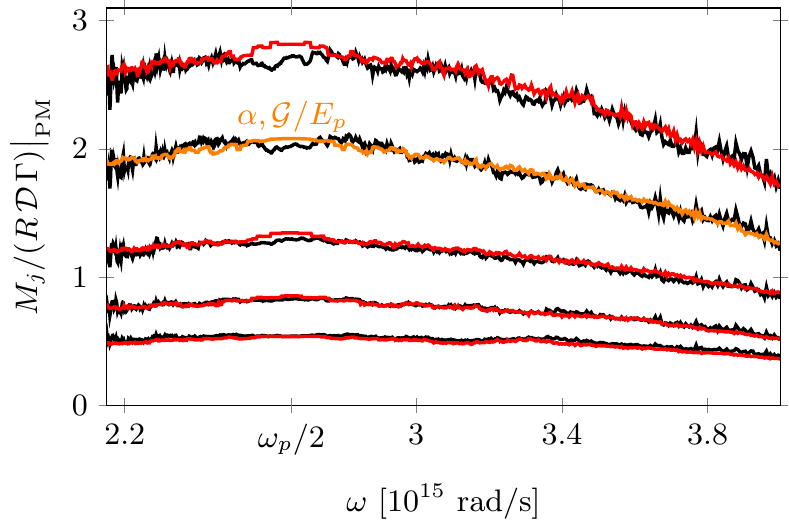}
\caption{
Maxima of densely packed high-gain spectra (right-hand side of equation~(\ref{eq:fit})) in black and their fit (left-hand side of equation~(\ref{eq:fit})) in red and orange, displayed in the frequency domain to highlight the distortion of the parabola. 
The fit parameters for the orange curve were obtained with the second-to-top measurement. 
To demonstrate their accuracy, we used the same fit parameters to draw the red curves.    
The fitting curves are noisy because fluctuations in the pump energy are taken into account. 
} 
\label{fig:high}
\end{figure}
 
In the absolute calibration measurement, we use a pump energy four times higher than in the spontaneous configuration.
We show the maxima of the spectra and the fit (orange curve) in Fig.~\ref{fig:high}.
The quantum efficiency at $\lambda = 2 \lambda_p$, extracted from fitting, is $\alpha = 0.42 \pm 0.04$, where the error is mainly due to our uncertainty in the pulse duration and transverse profile of the pump. 
Note that $\alpha$ includes all the losses in the optical setup, from the nonlinear crystal to the detector.
The estimated quantum efficiency of the experimental setup, based on the efficiency of each optical component, is $\alpha = 0.38 \pm 0.07$.
The largest source of loss is the diffraction grating of the spectrometer, with an efficiency of 60\% at $2\lambda_p$, as reported by the manufacturer.
In addition, we tested the consistency of the fit parameters by repeating the measurement with other pump energies. 
Using the previously obtained value of $\alpha$, and estimating the gain from $\mathcal{G}/E_p$ and a new measurement of $E_p$, we obtain the red curves (Fig.~\ref{fig:high}), which also show excellent agreement with experimental data. 

We note that equation~(\ref{eq:hgpm}) is based on a theoretical description where the pump is a monochromatic plane wave. 
The validity of this model in the context of a pulsed laser has been discussed~\cite{dayan2007theory} and verified experimentally by looking at the exponential increase in the number of photons with the pump power~\cite{agafonov2010two,allevi2014coherence,brida2009systematic, perez2014bright,spasibko2012spectral}.   
To our knowledge, the results presented in Fig.~\ref{fig:high} are the first experimental demonstrations of the distortion of the phase-matched spectral shape of light generated by a pulsed laser for increasing gain, and as such provide additional support for this description of PDC.

In contrast to the relative calibration, the absolute calibration using high-gain PDC cannot be straightforwardly generalized to arbitrary pump beams.
Corrections to the model could be implemented, for instance by taking into account the spatial profile and frequency spectrum of the pump as well as the frequency dependence of $\mathcal{G}$. 
However, our results demonstrate that even without a more sophisticated treatment, which would require the determination of many additional laboratory parameters and solving Heisenberg's equations of motion numerically,  we measure the quantum efficiency accurately. 

In summary, our work is based on the spontaneous generation of photon pairs triggered by the fluctuations of the joint vacuum field associated with the biphoton.
We demonstrated that the amplitude of vacuum fluctuations and its parametric amplification can serve as a primary radiation standard, and we used this insight to completely characterize a spectrometer.
As a first step, we used spontaneous PDC to correct for the instrument response function of a spectrum-measuring apparatus. 
Then, we retrieved the spectral quantum efficiency of the apparatus using the gain-dependent frequency spectrum of PDC in the high-gain regime.
The spectrum of our biphoton source is based on fundamental physical principles and is therefore comparable to Planck's law of radiation. 
In fact, the absolute calibration based on black-body radiation is also a two-step process, since the temperature must be accurately determined as well, often involving another measuring protocol such as filter radiometry~\cite{hollandt20055}.
In contrast to that, our two-step process is based solely on PDC and can therefore be performed with only one measuring apparatus, which could improve the accuracy, reliability, and reproducibility of metrological measurements.

\textbf{Methods.}
The third harmonic ($355\,$nm wavelength, $29.4\,$ps pulse duration, $50\,$Hz repetition rate, $100\,\mu$J pulse energy in the spontaneous regime, up to $500\, \mu$J in the high-gain regime) of a pulsed Nd:YAG laser is the pump for PDC from a nonlinear crystal ($\beta$-BBO, 3-mm thickness, type-I phase-matching, uncoated, cut for degenerate PDC) whose phase-matching frequencies are tuned using a motorized rotation mount. 
A set of dichroic mirrors remove the pump after the crystal.
The pump energy drift over time is monitored using a photodiode.
A concave mirror of focal length $200\,\textrm{mm}$ is used to bring the down-converted light to the far field, where a pinhole ($0.5\,\textrm{mm}$ diameter) selects a small solid angle.  
To ensure a fixed polarization, a broadband polarizing beam splitter is placed before the pinhole.   
A pair of lenses is used to image the pinhole onto the entrance slit of the spectrometer.
The spectrometer is an imaging spectrograph (Acton SP-2558) with a CCD camera (PIXIS:100BR\_eXcelon, pixels of size 20\,\textmu m\,$\times$\,20\,\textmu m).
Transverse binning is enabled, so that the signal at a certain wavelength is the sum of the photoelectron counts over all the pixels that correspond to that wavelength.
The integration time for each of the 411 spectra is $500\, \textrm{ms}$.
Each spectrum spans the range from $450\, \textrm{nm}$ to $900\, \textrm{nm}$. 
To cover this range, we need to repeat the acquisition for different angular positions of the grating (600 grooves per mm, 500-nm blaze). 
To reduce errors, we filter out the noise (rapidly fluctuating signal) in each spectrum with an algorithm based on fast-Fourier-transform.
The spectrometer is calibrated in wavelength using a neon-argon lamp along with Princeton Instruments Intellical system. 
The reference lamp (an LED-stack with a diffuser, Princeton Instruments) is introduced at the crystal plane.
Its spectrum is acquired using the same experimental settings. 

The datasets generated during and/or analysed during the current study are available from the corresponding author on reasonable request.

 \bibliography{calibration}

\begin{acknowledgments}
We thank Orad Reshef for valuable discussions.
This research was performed as part of a collaboration within the Max Planck-University of Ottawa Centre for Extreme and Quantum Photonics, whose support we gratefully acknowledge.
This work was supported by the Canada First Research Excellence Fund award on Transformative Quantum Technologies and by the Natural Sciences and Engineering Council of Canada.
R.F. acknowledges the financial support of the Banting postdoctoral fellowship of the NSERC and S.L. the financial support from Le Fonds de Recherche du Qu\'ebec Nature et Technologies (FRQNT). \\
\end{acknowledgments}
 
\noindent \textbf{Additional information.} Supplementary information is available in the online version
of the paper. 

\noindent \textbf{Author contributions.}
M.V.C. conceived the idea for relative calibration; S.L., E.G. and R.F. extended that idea to the absolute scheme.
S.L. and R.F designed the experiment, S.L. conducted the experiment and performed the data analysis. 
S.L., E.G. and R.F. wrote the manuscript. 
M.V.C. and R.W.B. supervised the project. 
All authors contributed to scientific discussions.

\noindent \textbf{Competing Interests.} 
S.L., M.V.C. and R.W.B., along with coinventors Mathieu Manceau et Gerd Leuchs, and the University of Ottawa and the Max Planck Institute for the Science of Light,  have a patent application (PCT/IB2017/056450) currently pending, about the relative calibration using parametric down-conversion. 
E.G. and R.F. declare that they have no competing financial interests. 

\end{document}


\title{Supplementary Material : A Primary Radiation Standard Based on Quantum Nonlinear Optics }

\author{Samuel~Lemieux}
\email[]{samzlemieux@gmail.com}
\affiliation{Department of Physics and Max Planck Centre for Extreme and Quantum Photonics, University of Ottawa, 25 Templeton Street, Ottawa, Ontario K1N 6N5, Canada}

\author{Enno~Giese}
\affiliation{Institut für Quantenphysik and Center for Integrated Quantum Science and Technology $\left(\text{IQ}^{\text{ST}}\right)$, Universität Ulm, Albert-Einstein-Allee 11, D-89081, Germany}

\author{Robert~Fickler}
\affiliation{Institute for Quantum Optics and Quantum Information (IQOQI), Austrian Academy of Sciences, Boltzmanngasse 3, 1090 Vienna, Austria}

\author{Maria~V.~Chekhova}
\affiliation{Max Planck Institute for the Science of Light, G.-Scharowsky Str.1/Bau 24, 91058 Erlangen, Germany}
\affiliation{Physics Department, Lomonosov Moscow State University, Moscow 119991, Russia}
\affiliation{University of Erlangen-Nuremberg, Staudtstrasse 7/B2, 91058 Erlangen, Germany}

\author{Robert~W.~Boyd}
\affiliation{Department of Physics and Max Planck Centre for Extreme and Quantum Photonics, University of Ottawa, 25 Templeton Street, Ottawa, Ontario K1N 6N5, Canada}
\affiliation{Institute of Optics, University of Rochester, Rochester, New York 14627, USA}

\date{\today}

\begin{abstract}
We discuss several relevant quantities for radiometry in a general manner, in particular the connection of the photon statistics of a quantized mode to the number of photons detected by a detector.
Further, we investigate the angular dependence of the intensity of down-converted light and the approximation used for angular mode selection by a pinhole and the wavelength dependence of the gain.
Also, we describe the experimental setup in detail and discuss details of the data analysis for both the spontaneous and the high-gain regime of parametric down-conversion.  
We finally prove that the low-gain experiments have been performed in the spontaneous regime.   
\end{abstract}

\maketitle
\normalsize

\section{Radiometry}
Since the quantization of the electric field is usually performed in plane-wave modes denoted by a wave vector $\vec{k}$, we express general radiometric quantities through the photon number per plane wave mode $\mathcal{N}(\vec{k})$ of the field under consideration.
A detector cannot detect all of these modes, and hence the detected photon-number density in the quantization volume can be written as
\begin{equation}
 \varrho= \frac{1}{(2\pi)^3} \smashoperator{\int_{\text{detector}}} \textrm{d}^3 k\, \mathcal{N}(\vec{k})= \smashoperator{\int_{\Delta \lambda}} \textrm{d}\lambda \smashoperator{\int_{\Delta \Omega}} \textrm{d}\Omega \, \frac{1}{\lambda^4}\, \mathcal{N}(\vec{k}),
\end{equation}
where we used $\textrm{d}^3 k = k^2 \textrm{d}k \textrm{d}\Omega = (2\pi)^3\lambda^{-4} \textrm{d}\lambda \textrm{d}\Omega$ in the last step.
We neglect here the index of refraction of air and assume that the detector has a bandwidth of $\Delta \lambda$ and collects light from a solid angle $\Delta \Omega$.
In the following we introduce for a more convenient notation the Jacobian $\mathcal{D}(\lambda)= (2\pi)^3 \lambda^{-4}$, which is proportional to the mode density.
For a sufficiently small detector bandwidth around the wavelength $\lambda$ and a small solid angle around $\Omega$, we can perform the integration and find
\begin{equation}\label{e_rho_approx}
 \varrho(\lambda,  \Omega)\cong  \frac{1}{(2\pi)^3}[\Delta \lambda \Delta \Omega]\mathcal{D}(\lambda) \mathcal{N}(\vec{k}),
\end{equation}
where $\mathcal{N}(\vec{k})$ implicitly depends on $\lambda$ and $\Omega$ through the wave vector $\vec{k}$.
This quantity is closely related to the spectral radiance $\hbar \omega (2\pi)^{-3}c \mathcal{D}(\lambda) \mathcal{N}$, which is
the energy per units of time, area of the source, solid angle and bandwidth (in wavelength) of the detector~\cite{datla20051}.

To calculate the total number of photons that are detected, the density from eq.~\eqref{e_rho_approx} has to be integrated over the volume of the source,
\begin{equation}
\label{eq:N_lambda}
N(\lambda,\Omega)= \smashoperator{\int_{\text{source}}} \textrm{d}^3 r\, \varrho(\lambda,  \Omega) \cong (2\pi)^{-3} [A_s c\tau_s][\Delta \lambda \Delta \Omega]\mathcal{D}\,\mathcal{N}
\end{equation}
where in the last step we assumed that the source has a surface area of $A_s$ and emits light for a time duration $\tau_s$.

We have not yet specified the photon distribution per plane wave mode $\mathcal{N}$.
We do that in the next section and show that the assumption of a small solid angle as well as a small bandwidth of the detector is justified.

\section{Angular distribution of spontaneous PDC}

The photon statistics per plane wave mode $\mathcal{N}$ for spontaneous PDC of a bulk crystal of length $L$ with a nonlinearity $\chi^{(2)}$ and illuminated by a plane wave pump with a field amplitude $E_p$ is
\begin{equation}\label{e_statistics}
\mathcal{N} = c^{-2}\,\left(L \chi^{(2)} E_p\right)^2\frac{\omega_s}{n_s(\omega_s)} \frac{\omega_i}{n_i(\omega_i)}\, \textrm{sinc}^2 ( \Delta \kappa L/2 ),
\end{equation}
where $\omega$ and $\omega_i$ are the frequencies of the signal and idler photons, and $n_s$ and $n_i$ their respective indices of refraction.
The longitudinal wave vector mismatch $\Delta \kappa = k_p-\kappa_s-\kappa_i$ is the difference of the wave vector $k_p$ of the pump and the longitudinal wave vectors $\kappa_{s,i}\equiv \sqrt{k_{s,i} - \vec{q}_{s,i}}$.
Here, the signal and idler photons have the wave vectors $k_s$ and $k_i$ and the transverse wave vectors $\vec{q}_s$ and $\vec{q}_i$.
Note that $k_j = \omega_j n_j / c$, with $c$ the speed of light and $\omega_j$ the frequency of the signal, idler and pump fields with $j=s,i,p$.

With this notation, we find the expression
\begin{equation}
\Delta \kappa =  k_p - k_s \left(\kappa_s/k_s + \sqrt{(k_i/k_s)^2 - (\vec{q}_i/k_s)^2} \right), 
\end{equation}
for the longitudinal wave vector mismatch.
Since we assume in \eqref{e_statistics} a plane wave and monochromatic pump, we have due to energy conservation $\omega_i=\omega_p-\omega_s$ and due to momentum conservation $\vec{q}_s = -\vec{q}_i$.
Hence, our expression depends only on $\omega_s$ and $\vec{q}_s$, which we can link to quantities of the detected field, which are written without a subscript. 
We find the connection $\omega_s = 2\pi c/\lambda$ when we express every quantity by the detected wavelength $\lambda$.
Moreover, introducing spherical coordinates, we can define the polar angle $\theta$ of the detected field and have $\cos \theta = \kappa_s/k_s$ and $\sin \theta = \vec{q}_s/k_s$.
The longitudinal wave vector mismatch
\begin{equation}
\Delta \kappa =  k_p - k_s \left(\cos \theta + \sqrt{(k_i/k_s)^2 - \sin^2 \theta} \right)
\end{equation}
therefore depends only on $\lambda$ and $\theta$, as does equation~\eqref{e_statistics}.

In eq.~\eqref{e_rho_approx} we approximated the integral of $\mathcal{N}$ over $\text{d}\lambda$ and $\text{d}\Omega = \sin \theta\text{d}\theta \text{d}\phi$ by just multiplying the integration intervals.
This is of course only valid if $\mathcal{N}$ depends weakly on both $\lambda$ and $\theta$ over the range of interest.

\begin{figure}[htb]
\includegraphics{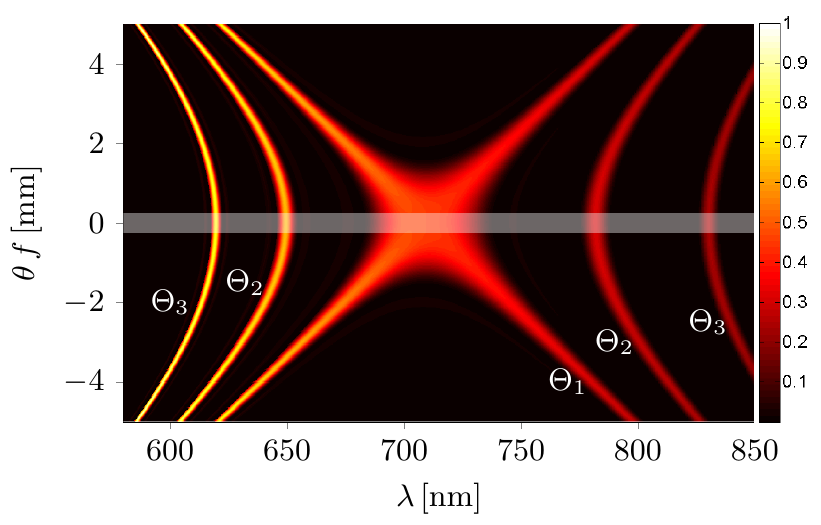}
\caption{
Numerically generated spectrum of spontaneous PDC, plotting $ \mathcal{D}(\lambda) \mathcal{N}$, for three different crystal tilt angles $\Theta$, with $\Theta_1$ corresponding to degenerate phase-matching in the emission angle $\theta = 0$.
While $\mathcal{N}$ is the spectral density in the $k$-space, $\mathcal{D}(\lambda) \mathcal{N}$ corresponds to the spectral density in the angular-wavelength representation, which is the measurement basis of our spectrometer. 
The vertical axis is represented in terms of the position in the far-field, using a concave mirror of focal length $f = 200\,\textrm{mm}$.
The semi-transparent white strip is the angular filtering of a pinhole of size $0.5\,\textrm{mm}$ positioned around $\theta=0$. 
  }
\label{fig:xcurve}
\end{figure}

In the experiment we place a pinhole in the far field of the spontaneous PDC light to filter a small range of angles.
We show in the density plot of Fig.~\ref{fig:xcurve} the product $\mathcal{D}(\lambda)\mathcal{N}$ as a function of $\theta$ and $\lambda$ and mark the size of our pinhole by a semi-transparent white strip.
This numerical result is based on the Sellmeyer equations of the three fields for BBO~{\cite{eimerl1987optical}.
We further assume that $\mathcal{G}$ is constant in the wavelength range of interest,  and we justify this assumption in the next section.
We work close to collinear propagation, with $\theta \approx 0$, where the function $\mathcal{D}(\lambda)\mathcal{N}$ does not vary significantly across the pinhole area so that we can perform the integration by just multiplying with the solid angle.
Similarly, the size of a pixel corresponds roughly to a bandwidth of $0.063$\,nm. 
On this scale, the function $\mathcal{N}$ does not change significantly.
Hence, our approximation in eq.~\eqref{e_rho_approx} is valid for our setup.

Of course, an integration of the pinhole angle can be performed as well to obtain an even more accurate result, but at some point the contribution of other crystal properties such as its length $L$ as well as the dispersion relations of all the light fields will dominate. 
In the spirit of an easy-to-implement calibration technique, we refrained from this more complex analysis but emphasize that it is possible.
In a similar manner, one could include both the frequency as well as the angular profile of the pump in eq.~\eqref{e_statistics}.
However, on axis this would not lead to a different result and our plane wave and monochromatic assumption is well-justified for our laser system.

\section{Wavelength dependence of gain}
In the main body of our article, we assumed that the wavelength dependence of the gain function
\begin{equation}\label{eq:gain}
\mathcal{G} = c^{-1} L \chi^{(2)} E_p / \sqrt{n_s n_i}
\end{equation}
can be neglected.
In this section, we investigate different effects that could contribute to the wavelength dependence in our experiment and demonstrate that they do not vary much across the spectral region of interest.
In addition to the linear dispersion ($n_s(\lambda)$ and $n_i(\lambda)$) as well as the nonlinear dispersion $\chi^{(2)}(\omega_p, \omega_s,\omega_i)$, obvious from eq.~(\ref{eq:gain}), other contributions arise from tilting the angle of the crystal to scan different phase-matching conditions.
By tilting the crystal, the Fresnel coefficients vary (for the pump or for the down-converted light) and the effective length $L$ of the nonlinear crystal (defined as the length of propagation of the pump inside the crystal) changes. 
The different Fresnel coefficients change the intensity of the pump inside the crystal, as well as how much of the down-converted light couples out of the crystal.
Using the Sellmeyer equations for BBO~\cite{eimerl1987optical} and Miller's rule~\cite{BOYD} (relating the first order and second order susceptibilities), we estimate the impact of those contributions, and show our results in Fig.~\ref{fig:cont}.
The largest deviations are attributed to the dispersion in the nonlinear susceptibility $\chi^{(2)}$  and to the change in the effective length of the nonlinear crystal upon tilting it.
However, over a spectral range of 300\,nm around degeneracy, the gain function $\mathcal{G}$ does not vary by more than 1\%.

\begin{figure}[htb] \centering
\includegraphics{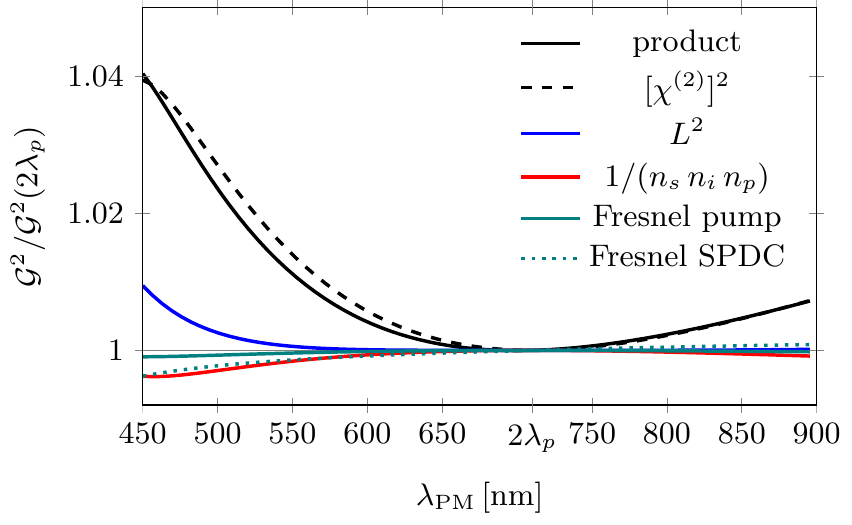}
    \caption{
Wavelength dependence of $\mathcal{G}^2$, normalized to its value at the degenerate frequency.
Each phase-matched wavelength $\lambda_\text{PM}$ corresponds to a set of refractive indices $n_s$, $n_i$ and $n_p$ (for the pump) that satisfy the phase-matching condition, occurring at a certain tilt of the nonlinear crystal.
The refractive indices also influence the nonlinearity $\chi^{(2)}$, appearing as well in eq.~\eqref{eq:gain}, through Miller's rule~\cite{BOYD}.
At a given energy per pulse, the electric field amplitude of the pump scales with $n_p^{-1}$. 
We combined this contribution with $1/(n_s n_i)$, which is explicit in the expression for $\mathcal{G}^2$.  
The varying angle between the pump propagation direction and the crystal leads to a different effective length $L$ of the crystal.
The Fresnel coefficients depend on the incidence angle and on the refractive indices: at the entrance facet the coefficients change how much of the pump $E_p$ is transmitted into the nonlinear medium; the exit facet changes the amount down-converted light that couples out.
The solid black line is the product all these effects.
}
\label{fig:cont}
\end{figure}

\begin{figure}[htb]
\includegraphics{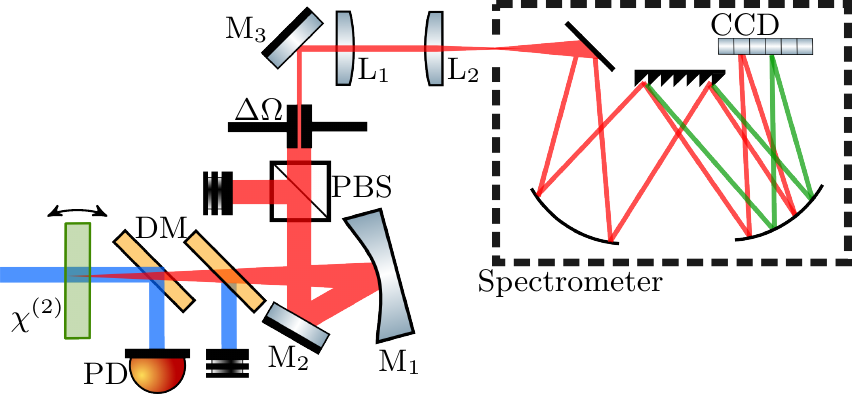}
\caption{
Experimental setup.
The down-converted light from the nonlinear crystal $\chi^{(2)}$ is angularly filtered in the far field by the pinhole $\Delta \Omega$. 
The plane of the pinhole is then imaged onto the entrance of the imaging spectrograph, which brings the grating-diffracted light onto the CCD camera.  
  }
\label{fig:setup}
\end{figure}

\section{Experimental setup}

The third harmonic~($355\,$nm wavelength, $29.4\,$ps pulse duration, $50\,$Hz repetition rate) of a pulsed Nd:YAG laser is prepared to serve as the pump for PDC: a pair of dispersive prisms suppresses the spurious frequencies from the laser; a half-wave plate and $\alpha$-BBO Glan-Laser polarizer set the polarization; a pair of lenses (focal lengths $f_1 = 300\,\textrm{mm}$ and $f_2 = 100\,\textrm{mm}$, separated by the distance $f_1 + f_2$) bring the diameter of the laser beam down to approximately $0.6\,$mm; a pinhole of size 100\,\textmu m is introduced between the lenses at the beam focus (distance $f_1$ from the first lens) to spatially filter the beam; and another $\alpha$-BBO Glan-Laser polarizer confirms the polarization of the beam.

The remainder of the experimental setup is shown in Fig.~\ref{fig:setup}. 
Parametric down-conversion is generated from the interaction of the pump beam with a nonlinear crystal $\chi^{(2)}$ ($\beta$-BBO, 3-mm thickness, type-I phase-matching, uncoated, cut for degenerate PDC with a 355-nm pump) on a motorized rotation mount. 
The wavelengths that satisfy the phase-matching condition are tuned by varying the angle between the optic axis of the crystal and the wavevector of the pump.
Two dichroic mirrors (DM) suppress the pump after the crystal and reflect the pump light onto a photodiode (PD) to monitor its intensity. 
A concave mirror (M$_1$) of focal length $200\,\textrm{mm}$ is used to bring the down-converted light to the far field, where a pinhole $\Delta \Omega$ ($0.5\,\textrm{mm}$ diameter) selects a small solid angle.  
A broadband polarizing beam spliter PBS placed before the iris is set to transmit the polarization of the down-converted light.   
A pair of lenses of (L$_1$) and (L$_2$) of focal lengths $200\,\textrm{mm}$ and $150\,\textrm{mm}$ are used to image the iris onto the entrance slit (1-$\textrm{mm}$ wide) of the spectrometer, with a magnification of $4/3$.
The spectrometer is an imaging spectrograph (Acton SP-2558) with a CCD camera (PIXIS:100BR\_eXcelon, $1340 \times 100$ pixels of size 20\,\textmu m$ \times $20\,\textmu m).  
The integration time for each spectrum is $500\, \textrm{ms}$.
Transverse hardware binning~(summing the photoelectron count for the 100 transverse pixels) is enabled.
Each spectrum spans the range from $450\, \textrm{nm}$ to $900\, \textrm{nm}$. 
To cover this range, we need to repeat the acquisition for different angular positions of the grating (600 grooves per mm, 500-nm blaze). 
The experiment is automated: after each acquisition by the spectrometer, the motorized holder rotates the crystal through an angle of about $0.01\degree$, up to a total change of approximately 8\degree. 
The pump energy measured at the photo-diode is recorded for each position of the crystal. 
The wavelength of the spectrometer is calibrated using a neon-argon lamp along with Princeton Instruments Intellical system. 
After the experiment a reference lamp is introduced at the crystal plane.
Its spectrum is acquired using the same experimental settings.
The reference response function that we use to verify our method is obtained by comparing the measured spectrum of a calibration lamp (LED-stack with a diffuser for relative intensity calibration) and a reference spectrum provided by Princeton Instruments.

\section{Details on the data analysis}

Our calibration method relies on the comparison of the measured phase-matched number of counts $M(\lambda_\textrm{PM})$ to the expected number of phase-matched photons $N(\lambda_\textrm{PM})$. 
We therefore acquire a large number of spectra $M_j$ corresponding to different phase-matching conditions over a broad spectral range.
However, the peak number of counts in a measured spectrum does not correspond, in general, to $M(\lambda_\textrm{PM})$. 
Instead, we can extract the response function from the properties of $\mathcal{N}$. 
From the main text, we know that 
\begin{equation}
\mathcal{N} \sim  \omega(\omega_p - \omega) \, \mathrm{sinc}^2 ( \Delta \kappa  L/2 ) \leq \omega (\omega_p - \omega), 
\end{equation}
where the inequality becomes equality only for phase matching $\Delta \kappa = 0$. 
We denote the wavelength of phase matching with $\lambda_\text{PM}$.
With eq.~(5) from the main text we find the inequality
\begin{equation}
\label{eq:Rgreater}
R(\lambda)\geq R(\lambda) \operatorname{sinc}^2 \frac{\Delta \kappa L}{2} \propto \frac{M_j(\lambda)}{\mathcal{D}(\lambda) \omega (\omega_p-\omega)}
\end{equation}
with an equality sign for $\lambda=\lambda_\text{PM}$. 
If we approximate the phase matching function by a Gaussian, i.e.,
\begin{equation}
\textrm{sinc}^2(\Delta \kappa L /2 ) \propto \textrm{exp}[ -(\lambda - \lambda_{\textrm{PM}})^2/(2\sigma_\lambda ^2)],
\end{equation}
it is easy to show that the peak of the product $R(\lambda)\operatorname{sinc}^2 (\Delta \kappa L/2)$ shifts to the wavelength
\begin{equation}
\tilde{\lambda}  = \lambda_{\textrm{PM}} + \frac{1 }{R}\frac{\textrm{d}R }{\textrm{d}\lambda}\bigg\rvert_{\tilde{\lambda}} \sigma_\lambda^2.
\label{eq:shift}
\end{equation}
Hence, the shift between phase-matched wavelength and peak increases, the steeper the slope of the response function or the wider the peak is.
Since the response function  is not known but is the result of the calibration procedure, eq.~\eqref{eq:shift} cannot be used to determine the phase-matching wavelength.
However, eq.~\eqref{eq:Rgreater} directly gives a method to determine the response function despite the shift:
when we acquire a large number of spectra $M_j$, each with a slightly varying $\lambda_\text{PM}$, the amplitude of $M_j/[\mathcal{D}\omega (\omega_p-\omega)]$ at one particular wavelength is the largest if the wavelength corresponds to $\lambda_\text{PM}$.
Hence, we obtain the response function from
\begin{equation}
R(\lambda) = \underset{j}{\max} \left[ \frac{M_j(\lambda)}{\mathcal{D}(\lambda) \omega (\omega_p -\omega)} \right] \bigg/  \underset{j}{\textrm{max}} \left[ \frac{4M_j(2 \lambda_p)}{\mathcal{D}(2 \lambda_p) \, \omega_p^2} \right],
\label{eq:max}
\end{equation}
where we normalize the response function to unity at the degenerate wavelength $\lambda = 2 \lambda_p$.
To reduce errors in the analysis according to eq.~\eqref{eq:max}, we suppress for each $M_j(\lambda)$ spectrum the high-frequency content, filtered out via a fast-Fourier-transform procedure. 

A similar idea can be used for absolute calibration.
For an arbitrary $\mathcal{G}$, the photon distribution per plane-wave mode assuming a monochromatic plane wave pump can be written as~\cite{klyshko1989photons}
\begin{equation}\label{eq:N_HG}
\mathcal{N}^\text{(HG)}=\frac{\mathcal{G}^2 \mathcal{Q}^2}{\mathcal{G}^2 \mathcal{Q}^2-(\Delta \kappa L/2)^2} \sinh^2 \sqrt{\mathcal{G}^2 \mathcal{Q}^2-(\Delta \kappa L/2)^2},
\end{equation}
where $\mathcal{Q}^2\equiv \omega (\omega_p-\omega)$, and the superscript $\text{(HG)}$ highlights that we are using this equation to describe the high-gain regime of PDC.
Since the maximum of this function occurs for phase matching ($\Delta \kappa = 0$), we find
\begin{equation}
\label{eq:N_PM}
\mathcal{N}^\text{(HG)} \leq \sinh^2 (\mathcal{G}\mathcal{Q}) \equiv \mathcal{N}^\text{(HG)}_\text{PM}
\end{equation}
where we defined the phase-matched photon distribution $\mathcal{N}^\text{(HG)}_\text{PM}$ that has the well-known hyperbolic form of parametric amplification and is used in the main body of our article.
Note further that for $\mathcal{GQ}\ll 1$ we recover the low-gain result.

The quantum efficiency at the degenerate wavelength $\alpha = \eta (2\lambda_p)$ is
\begin{equation}
\alpha = M_j(\lambda)/[R(\lambda)N(\lambda)]
\end{equation}
with the definitions from the main body of the article.
With that, we find from eq.~\eqref{eq:N_PM} and with the help of eq.~\eqref{eq:N_lambda} the inequality
\begin{equation}
\alpha \sinh^2 \mathcal{G}\mathcal{Q} \geq M_j(\lambda)/[R(\lambda)\mathcal{D}(\lambda)\Delta \Omega\Delta \lambda A_s c \tau_s],
\end{equation}
where again the equal sign is valid for $\lambda = \lambda_\text{PM}$.
Hence, we find, similarly to the low-gain method,
\begin{equation}
\alpha \sinh^2  \mathcal{G}\mathcal{Q}= \underset{j}{\max}\left[ \frac{M_j(\lambda)}{R(\lambda)\mathcal{D}(\lambda)\Delta \Omega\Delta \lambda A_s c \tau_s}\right]
\end{equation}
as an exact equality if the spectra are sufficiently dense.
Taking the maximum of all recorded spectra, each one of them divided by $R(\lambda)\mathcal{D}(\lambda)$ and a numerical factor that depends on laboratory parameters (spatial dimensions and bandwidths), we can fit the data to the function $\alpha \sinh^2  \mathcal{G}\mathcal{Q}$ with two fitting parameters $\alpha$ and $\mathcal{G}$.
Note that we do not need to measure the exponential increase of the generated photons with increasing pump intensity, but determine both parameters from the distortion of the \emph{spectral shape} of the maximum of all spectra.
With this fitting procedure, one can not only determine the quantum efficiency $\eta(\lambda) = \alpha R(\lambda)$, but also the gain $\mathcal{G}$.

Even though we do not use the exponential increase with the pump power for our calibration method, we still record the intensity while scanning different phase matching functions.
We do this to correct for drifts and fluctuations during the course of one measurement.
We are then able to perform the fitting procedure using $\mathcal{G}/E_j$, where $E_j$ is the pump field amplitude during measurement corresponding to the the $j$th phase-matching condition.

The $\alpha$ obtained using our method for absolute calibration is compared to an estimated quantum efficiency based on the properties of each optical component in the experimental setup, listed in table~\ref{table:1}. 
The efficiency of uncoated components is estimated from the Fresnel coefficients, while the efficiency of coated components is taken from the manufacturers.

\begin{table} [htb]
\caption{Contribution of each optical component to the total quantum efficiency of the experimental setup. 
The parentheses denote the number of components.
The total efficiency is obtained by multiplying all the contributions and propagating the uncertainties accordingly.   
 }
\centering
\begin{tabular}{lr}
\hline
Optical component \quad & Efficiency \\
\hline
Crystal output facet (1)    &  $0.94 \pm 0.01$      \\
Dichroic mirror (2)    &  $0.95 \pm 0.01$      \\
Dielectric mirrors (6) \quad \quad  &  $0.99 \pm 0.01$      \\
Polarizing beam splitter (1) \quad \quad    &  $0.98 \pm 0.01$      \\
Uncoated lens (2)   &  $0.92 \pm 0.01$      \\
Diffraction grating (1)  &  $0.60 \pm 0.02$      \\
Spectrometer camera (1)  &  $0.95 \pm 0.02$      \\
\hline
Total    &     $0.38 \pm 0.07$  \\
\hline
\end{tabular}
\label{table:1}
\end{table}

\section{Spontaneous regime of parametric down-conversion}

\begin{figure}[htb]
\includegraphics{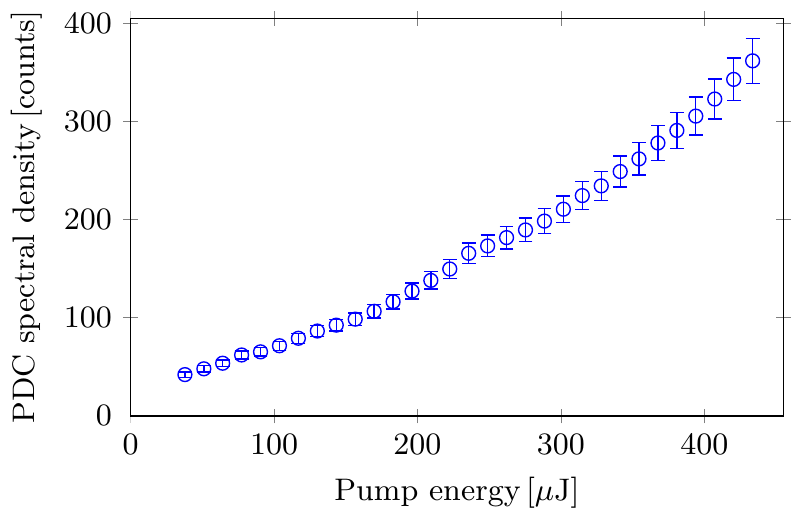}
\caption{
PDC spectral density as a function of the pump energy per pulse. The number of counts was extracted at the phase-matched wavelength $\lambda = 690\,\textrm{nm}$. The error bars are obtained from the amplitude of the noise in the spectrum. The acquisition time is 5 seconds, for a total of 250 pulses. 
  }
\label{fig:lin}
\end{figure}

As shown in the section above in eq.~\eqref{eq:N_HG}, the photon-number distribution grows exponentially with the intensity of the pump.
In the low-gain regime, where the photon pairs are generated spontaneously, the number of photons grows linearly with the intensity which can be seen from the expansion
\begin{equation}
\mathcal{N}_\text{PM}^\text{(HG)} = \sinh^2 \mathcal{GQ} \cong \mathcal{G}^2\mathcal{Q}^2 = \mathcal{G}^2 \omega(\omega_p-\omega) = \mathcal{N}_\text{PM} ,
\end{equation}
where $\mathcal{N}_\text{PM}$ is the low-gain photon distribution for phase matching.
To obtain the response function $R(\lambda)$, we do not need to know the exact value of $\mathcal{G}$ but rely on the fact that the first-order expansion above is valid.
Note that $\mathcal{G}^2$ is proportional to the intensity of the pump $I_p$~\cite{klyshko1989photons}.
To verify that we work in the spontaneous regime of PDC, we measure the number of counts for a single wavelength and increase the pump intensity.
The results are shown in Fig.~\ref{fig:lin}.
We see, that we are well in the linear regime up to roughly 150\,\textmu J.
We performed the relative calibration experiment at a pump intensity of 100\,\textmu J, while the high-gain part of the experiment used a more intense pump, around 200\,\textmu J and higher.

 \bibliography{supp}